\pdfoutput=1 
\documentclass[runningheads]{llncs}
\usepackage[dvipsnames]{xcolor}

\newcommand{\Lagr}{\mathcal{L}}
\usepackage{amsmath}

\usepackage{graphicx}
 \title{Differentiable Deconvolution for Improved Stroke Perfusion Analysis}
\author{Ezequiel de la Rosa\inst{1,2} \and
David Robben\inst{1,3,4} \and
Diana M. Sima\inst{1} \and
Jan S. Kirschke\inst{5} \and
Bjoern Menze\inst{2}}

\authorrunning{E. de la Rosa et al.}

\institute{ico\textbf{metrix}, Leuven, Belgium\\
\email{\{ezequiel.delarosa, david.robben, diana.sima\}@icometrix.com}\\
\and
Department of Computer Science, Technical University of Munich, Munich, Germany\\
\email{bjoern.menze@tum.de}\\
 \and
Medical Imaging Research Center (MIRC), KU Leuven, Leuven, Belgium
 \and
Department of Electrical Engineering, ESAT/PSI, KU Leuven, Leuven, Belgium
 \and
 Neuroradiology, School of Medicine, Technical University of Munich, Munich, Germany\\
\email{jan.kirschke@tum.de}}

\begin{document}
\maketitle              
% typeset the header of the contribution
%
\begin{abstract}
Perfusion imaging is the current gold standard for acute ischemic stroke analysis. It allows quantification of the salvageable and non-salvageable tissue regions (penumbra and core areas respectively). In clinical settings, the singular value decomposition (SVD) deconvolution is one of the most accepted and used approaches for generating interpretable and physically meaningful maps. Though this method has been widely validated in experimental and clinical settings, it might produce suboptimal results because the chosen inputs to the model cannot guarantee optimal performance. For the most critical input, the arterial input function (AIF), it is still controversial how and where it should be chosen even though the method is very sensitive to this input. In this work we propose an AIF selection approach that is optimized for maximal core lesion segmentation performance. The AIF is regressed by a neural network optimized through a differentiable SVD deconvolution, aiming to maximize core lesion segmentation agreement with ground truth data. To our knowledge, this is the first work exploiting a differentiable deconvolution model with neural networks. We show that our approach is able to generate AIFs without any manual annotation, and hence avoiding manual rater's influences. The method achieves manual expert performance in the ISLES18 dataset. We conclude that the methodology opens new possibilities for improving perfusion imaging quantification with deep neural networks.    

\end{abstract}

\section{Introduction}
Perfusion imaging techniques are the clinical standard for acute ischemic stroke lesion assessment. They acquire images of the passage of a contrast agent bolus through the brain tissue. Since the perfusion series are not directly clinically interpretable, they require the computation of physically meaningful parameter maps. Although different approaches may be used for their computation (e.g. compartmental models,  which are mainly used over long acquisition time perfusion MRI), the preferred technique in perfusion CT analysis is the singular value decomposition (SVD) deconvolution \cite{lin2016whole,vagal2019automated}. The technique has been well validated in experimental \cite{murphy2007serial} and clinical \cite{albers2016ischemic} settings and is widely implemented in perfusion CT software \cite{fieselmann2011deconvolution,vagal2019automated}. Cerebral blood flow (CBF) and time to the maximum residue function (Tmax) are typically used maps, though cerebral blood volume and time-to-peak maps are often considered as well. Parameter maps are critical for treatment decision making. They allow assessing the salvageable $penumbra$ and irreversible $core$ necrotic lesions, and hence determining if reperfusion techniques may reduce the disease damage severity.

The SVD deconvolution method requires as input an arterial input function (AIF), defined as the concentration time-curve inside an artery feeding the tissue under study. In practice, the AIF is mostly selected by a physician, a demanding, highly variable and poorly reproducible process. Its correct selection is the cornerstone for generating accurate maps, as has been shown that minimal changes in its location and/or shape may strongly impact the deconvolution process \cite{mlynash2005automated}. Although several works studied how and where the AIF should be chosen \cite{calamante2013arterial}, the subject is still very controversial. Thus, AIF selection is suboptimal, since we do not know which function will maximize the deconvolution performance. Besides, the AIF's concept is defined based on the SVD-deconvolution theoretical model, which relies on several assumptions violated in clinical practice. For instance, limited voxel resolution, partial volume effect, time-curve delays, noise and other confounders are typically limiting the model's performance. Consequently, it is not straightforward to define which AIFs are the best to use in practice.

In this work we use neural networks to generate the AIF, aiming to find the $best$ AIFs in core lesion segmentation terms.  Through experiments on the ISLES18 database we show that the method is able to learn from scratch to generate AIFs that maximize the segmentation agreement with manually delineated ground truth. Thus, the AIFs are learned without any expert annotations of the AIFs themselves, hence avoiding potential rater's bias. We show, as well, that the approach is able to yield manual expert performance in the ISLES18 database.

\section{Method}
\subsection{Differentiable Deconvolution} 
\begin{figure}[b]
\includegraphics[width=\textwidth]{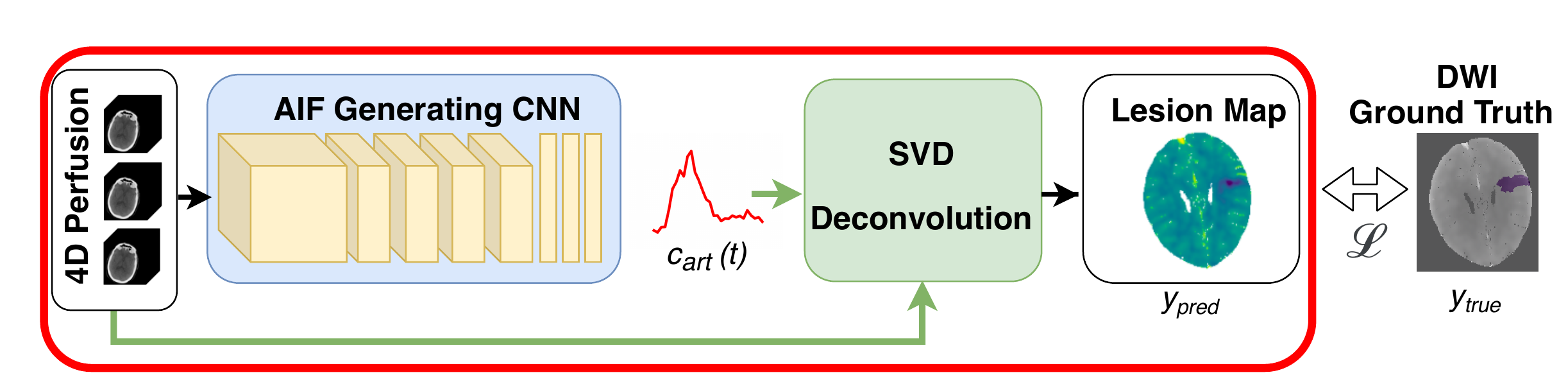}
\caption{Differentiable SVD deconvolution  pipeline.} \label{fig1}
\end{figure}

We propose the optimizable framework of Fig. \ref{fig1} which generates the $best$ AIF for SVD deconvolution. We define the $best$ AIF as the one yielding the highest agreement between the estimated core lesion and the ground truth core lesion. The input to the framework is 4D perfusion data and the output is a lesion map. It consists of a CNN that generates the AIF $c_{art}(t)$. The generated AIF together with the 4D input perfusion series pass through a differentiable SVD deconvolution block, which outputs relative CBF (rCBF) maps after image deconvolution. Finally, rCBF is transformed into the lesion probability map $y_{pred}$. The framework is end-to-end trainable, which means that gradients are backpropagated through all blocks including the SVD deconvolution, thus allowing to generate the best AIF candidate that maximizes the segmentation performance.  

\subsubsection{AIF Generating CNN}
Unlike most previous works that use unsupervised clustering \cite{mouridsen2006automatic,murase2001determination,peruzzo2011automatic,shi2014automatic,rausch2000analysis} or supervised segmentation \cite{fan2019automatic} approaches, we propose a regression CNN for obtaining an AIF $c_{art}(t)$ from the 4D perfusion series. The architecture is fairly straightforward: the only particularity is that the input size varies in the z-axis from 2 to 8 slices. Therefore, it consists of 3D convolutional layers followed by an average pooling over the z-axis. Subsequently, there are 2D convolutional layers and finally a fully connected layer with the same number of output neurons as the number of time-points in the perfusion data. This final 1D vector represents $c_{art}(t)$. After each convolution, average pooling and dropout \cite{srivastava2014dropout} are used. ReLU \cite{krizhevsky2012imagenet} activations are applied in all layers except for the final AIF output layer, where a linear activation is employed. For fitting 4D data into the CNN, we encode volume time-points as channels in the network.

\subsubsection{SVD Deconvolution}
\label{subsec:SVD-Deconvolution}
For generating interpretable and physically-meaningful maps, perfusion series are deconvolved using the well-validated delay-invariant SVD deconvolution. Our GPU implementation of the algorithm uses Volterra discretization \cite{sourbron2007deconvolution}. Differentiability of the SVD algorithm was studied in \cite{papadopoulo2000estimating,townsend2016differentiating}. Through analytical methods it was shown that the Jacobian of the SVD is computationally feasible \cite{papadopoulo2000estimating}. Even more, in \cite{ionescu2015training} a differentiable SVD was used in neural network layers for a variety of tasks including image segmentation. In this work, we take advantage of current deep learning libraries that support auto-differentiation and allow backpropagating SVD gradients. Given the fact that deconvolution is an ill-conditioned problem \cite{robben2018perfusion,fieselmann2011deconvolution}, we use a Tikhonov regularization scheme. The output of the deconvolution block of Fig. \ref{fig1} is rCBF. Although in clinical practice the Tmax and rCBF maps are typically used for identifying core lesions, we only consider rCBF since the differentiability of Tmax is not fully clear.

In a nutshell, the analysis of perfusion data can be described by the following convolution product:
\begin{equation}
c_{voi}(t_j)= \int_{0}^{t}  c_{art}(\tau)k(t_j-\tau)d\tau
\label{eq:deconv_product}
\end{equation}
where $c_{voi}$ is the agent concentration in the voxel under consideration, $c_{art}$ is the concentration curve measured in an artery feeding the volume (i.e., the AIF), and $k$ is the impulse response function that characterizes the tissue of interest \cite{sourbron2007deconvolution,fieselmann2011deconvolution}. Note that here $t=0$ is taken prior to the arrival of the contrast agent (such that $k(t)=0$ for $t<0$).
For solving Equation \ref{eq:deconv_product} (i.e., finding $k$) we rely on discretization methods, since in practice the measured AIF $c_{art}$ and the agent concentration in the tissue volume  $c_{voi}$ are discretized at specific time points. Considering these time points as $t_j = (j-1)\Delta t$ (for $j=1,...,N$), and assuming that $c_{art}(t)$ is negligible for $t>N\Delta t$, the discretization of Equation \ref{eq:deconv_product} can be approximated as:

\begin{equation}
c_{voi}(t_j)= \int_{0}^{t}  c_{art}(\tau)k(t_j-\tau)d\tau \approx \Delta t \sum_{i=1}^{N}c_{art}(t_i)k(t_{j-i+1})  
\end{equation}
which can be rewritten as a linear system:

\begin{equation}
    \textbf{c=Ak}
    \label{eq:linear}
\end{equation}
where \textbf{A} is the Volterra matrix with the following $A_{ij}$ elements~\cite{sourbron2007deconvolution}:

\begin{equation}
\begin{cases}
    A_{i0}= (2c_{art}(t_{i}) + c_{art}(t_{i-1}))/6 & (0 <i \leq N-1)\\  
    A_{ii}= (2c_{art}(t_{i}) + c_{art}(t_{i+1}))/6 & (0 <i \leq N-1) \\
    A_{ij}= \frac{2}{3}c_{art}(t_{i}) + \frac{c_{art}(t_{i-1})}{6}+ \frac{c_{art}(t_{i+1})}{6} &(1 <i \leq N-1, 0<j<i)  \\
    A_{ij}= 0 & \text{elsewhere}
\end{cases}
\end{equation}
We assume, without loss of generality, that $\Delta t = 1 s$ , which is usually the case in clinical practice. For the derivation of \textbf{A} and for a more in-depth understanding of discretization methods in perfusion imaging, the reader is referred to \cite{sourbron2007deconvolution}.

A classical way of solving Equation \ref{eq:linear} is by means of SVD as:

\begin{equation}
    \mathbf{\textbf{A}=\textbf{U}\Sigma V^{T}=\sum_{i=1}^{r}u_{i}}\sigma_{i}\mathbf{v_{i}^{T}}
\end{equation}
where $r=rank(\mathbf{A})$, $\mathbf{U=[u_1,...,u_r]}$ and $\mathbf{v=[v_1,...,v_r]}$ are the left and right singular vectors, respectively, and $\mathbf{\Sigma}=diag(\sigma_1,..., \sigma_r)$ is the diagonal matrix containing singular values in decreasing order.
The least squares solution of Eq.~\ref{eq:linear} for $\mathbf{k}$ is:

\begin{equation}
    \mathbf{k = \sum_{i=1}^{r} \frac{u_{i}^{T}c}{\sigma_{i}}v_i} 
    \label{eq:solution_svd}
\end{equation}
Nonetheless, in cases where $\mathbf{A}$ is ill-conditioned, Eq. \ref{eq:solution_svd} is not a suitable solution of the linear system since a small variability in $\mathbf{c}$ may generate very large variability in $\mathbf{k}$ \cite{fieselmann2011deconvolution}. Thus, regularization is required for having a stable result as:

\begin{equation}
    \mathbf{k_{\lambda} = \sum_{i=1}^{r} }
    \left(\textit{f}_{\lambda,i}\mathbf{\frac{u_{i}^{T}c } {\sigma_{i}}}\right) \mathbf{v_i }
\end{equation}
where $\textit{f}_{\lambda,i} = \frac{\sigma_i^2}{\sigma_i^2+\lambda^2}$ are Tikhonov regularization parameters with $\lambda=\lambda_{rel}\sigma_i$. The parameter $\lambda_{rel}$ should be chosen in the interval (0, 1). In our implementation we empirically set $\lambda_{rel}=0.3$.
Finally, it can be proven that the cerebral blood flow can be obtained as:
%\begin{equation}
 %   \text{CBF}= \frac{1}{\rho_{voi}}\max(k(t))
%\end{equation}
\begin{equation}
    \text{CBF}= \frac{1}{\rho_{voi}}\max(k(t_{j}))
\end{equation}
where $\rho_{voi}$ $[\frac{g}{ml}]$ stands for the mean tissue density in the voxel. 
For a mathematical demonstration of this statement the reader is referred to~\cite{fieselmann2011deconvolution}. Finally, following current clinical practice, CBF is normalized with mean healthy CBF values for obtaining a map in a subject-independent scale.  

\subsubsection{Ischemic Lesion Map}
Generated rCBF maps require some sort of transformation to obtain lesion probability maps $y_{pred}$ that can be compared with the binary ground truth masks. With this aim, we use sigmoid activations centered at rCBF = 0.38 for mapping the $y_{pred}$ probability values. This cutoff previously was found to be optimal for this dataset \cite{cereda2016benchmarking}. It is worth to mention that it is possible to allow the network to choose the best cutoff, but we preferred to keep a fixed threshold as mostly used in clinical practice. In such a way, our proposed method is directly comparable with the results of a manual AIF selection using the same cutoff value, assuring that differences in results are only driven by the choice of AIF.

\subsection{Implementation and Optimization}
The framework is implemented using TensorFlow and Keras, where we ensure effective gradient propagation by only using differentiable operations. Given the class imbalance between $healthy$ and $necrotic$ brain tissue, a soft-Dice loss function is used as follows:

\begin{equation}
    \Lagr = 1 - \frac{2\sum y_{true} y_{pred}}{\sum y_{true} + \sum y_{pred} }
\end{equation}
where $y_{pred}$ is the framework's output lesion map and $y_{true}$ the ground truth manually delineated lesion mask. 

Optimization is conducted using stochastic gradient descent with momentum, with a unitary batch size. In order to improve the network's learning stage and to overcome data limitations, two types of data augmentation are conducted. First, perfusion specific data augmentation \cite{robben2018perfusion} is implemented at an image level, which allows mimicking AIF bolus delay arrivals and AIF peak concentration scaling. Second, traditional segmentation data augmentation is used, including image rotation, translation, flipping and random Gaussian noise addition.   

\subsection{Data and Experiments}
The free and open ISLES18 database is used \cite{maier2017isles,kistler2013virtual}. It consists of 4D CT perfusion series with ground truth core lesion delineations obtained from diffusion weighted imaging (DWI). From the total amount of scans provided in the challenge (n = 156), only the training set (n = 94) includes ground truth data and hence was used for our experiments. The dataset is multi-scanner and multi-center, obtained from different institutions from the U.S. and Australia. All provided images are already motion-corrected, co-registered for matching CTP with DWI modalities, and spatio-temporally resampled (with 256x256 images and 1 volume/second).

To compare our proposed method against the current clinical approach, an expert provided manual AIF annotations for the entire ISLES18 training set. A single global AIF was selected per case, following recommendations found in the literature \cite{calamante2013arterial}. Moreover, our results are compared with the automatic AIF selection approach included in ico\textbf{brain cva} (ico\textbf{metrix}, Leuven, Belgium), an FDA cleared CTP analysis software package. In both cases, the CTP images are deconvolved with the chosen AIF. The same data preprocessing and deconvolution algorithm of section \ref{subsec:SVD-Deconvolution} are used. As such, any difference in results is only caused by the AIFs themselves. rCBF maps are generated and the core lesions are quantified. For all our experiments and methods a fixed threshold rCBF $=$ 0.38  is used for defining the core lesions \cite{cereda2016benchmarking}.

Since the dataset is already preprocessed, the preprocessing on our side is limited to a spatio-temporal smoothing before the CTP data is deconvolved. The method's performance is assessed through 5-fold cross-validation. Results are evaluated at parameter map and lesion segmentation levels. The discriminant power of the rCBF maps for differentiating healthy and necrotic tissue are assessed through the area under the ROC curve (AUC). The lesion segmentation is assessed by means of Dice and Jaccard indexes, 95\% Hausdorff distance and volumetric Bland-Altman analysis \cite{bland1986statistical}.

\section{Results}
Training of our model takes $\sim$ 2.5 hours on an Nvidia K80 GPU with 12 GB dedicated memory. During testing, the entire process of AIF selection, SVD deconvolution and lesion quantification takes $\sim$ 1.15 seconds per case. On the other hand, the expert annotation of an AIF takes around one minute per case.
\begin{figure}[t]
\begin{center}
\includegraphics[width=9cm]{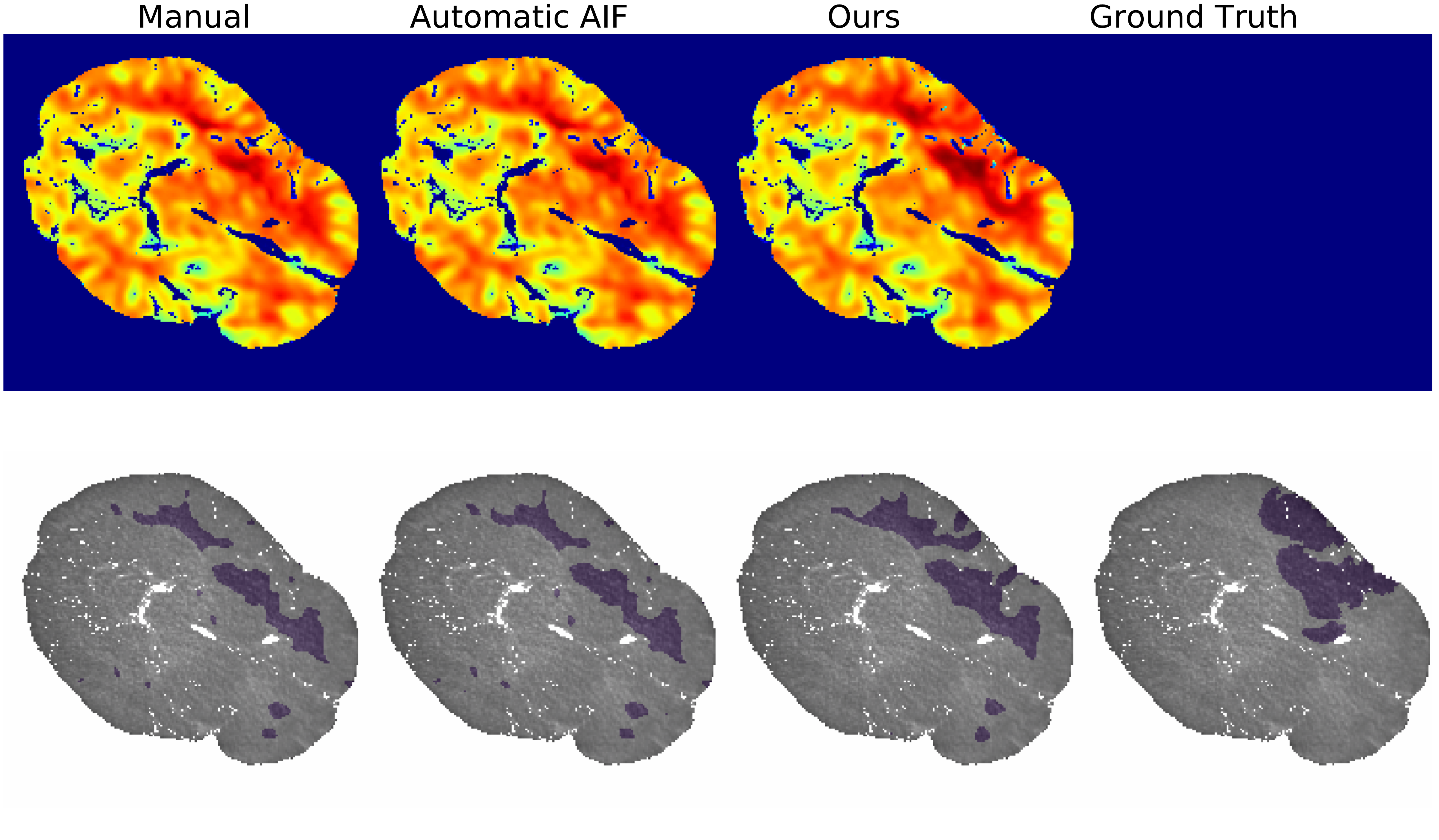}
\caption{Qualitative results. Top: rCBF maps; Bottom: Core-lesion segmentations.} \label{fig_quality}
\end{center}
\end{figure}

In Fig. \ref{fig_quality}, the resulting rCBF maps generated with the different methods are shown for an example case. The corresponding AIFs selected or generated by these methods are included in the supplementary material. The automatic algorithm yields results that are visually similar to the ones generated with the expert AIF selection. However, our generated map better matches the DWI ground truth lesions than both other approaches. When assessing the ROC AUC values of rCBF for discerning healthy and necrotic tissue, our proposed method shows better performance than the other methods (Table \ref{tab1}). In all segmentation metrics the methods achieve comparable performance, with our new method outperforming the others, but the differences are not statistically significant (paired t-test).

\begin{figure}[t]
\includegraphics[width=\textwidth]{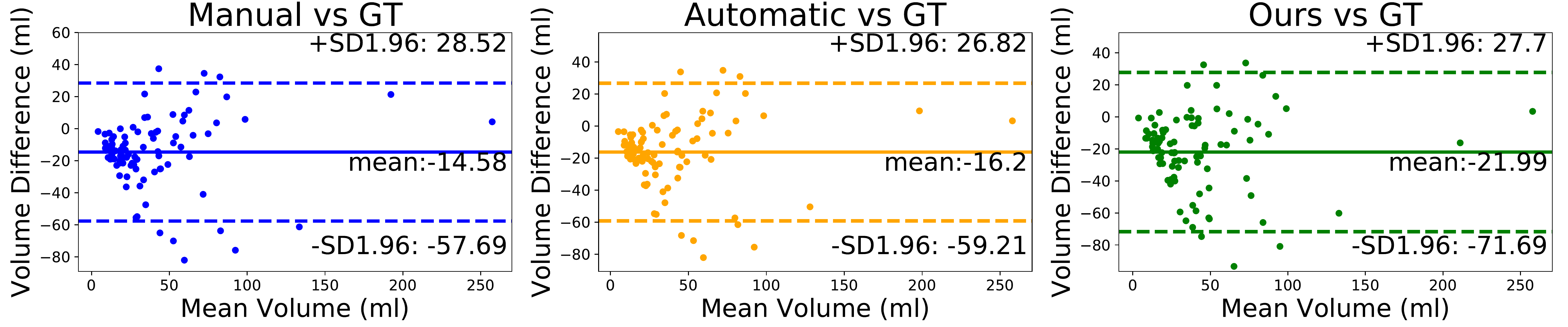}
\caption{Volumetric Bland-Altman plots. GT: Ground truth.} \label{fig_ba}
\end{figure}

\begin{table}[b]
\centering
\caption{Mean (standard deviation) of various segmentation metrics for the different methods. rCBF: relative cerebral blood flow; AUC: area under the ROC curve; HD: Hausdorff distance.}\label{tab1}
\begin{tabular}{|c|c|c|c|c|}
\hline
Method &  rCBF AUC & Dice & Jaccard & 95$\%$ HD (mm)\\
\hline
Expert & 0.856 & 0.353 (0.201) &0.233 (0.157)& 54.136 (19.449)\\
Automatic AIF &0.856  &0.351 (0.202) &0.232 (0.158) & 55.015 (18.790)\\
Ours& \textbf{0.868} & \textbf{0.359 (0.201)} & \textbf{0.238 (0.157)}& \textbf{53.747 (18.875)}\\
\hline
\end{tabular}
\end{table}
Fig. \ref{fig_ba} shows the lesion volume quantification performance in Bland-Altman plots. Volumetric overestimation is found for all methods when comparing with results reported by \cite{cereda2016benchmarking}, which shows better agreement with ground truth. This can be explained by their use of an extra Tmax criterion and due to their use of a modified ground truth, that excluded tissue with a low Tmax. The manual rater and the automatic AIF software yield comparable results, with the expert annotations having the best agreement with the ground truth (p-value non-statistically significant between these methods, Mann-Whitney U test). Our approach shows a larger volumetric bias than these methods, that was only statistically significant when compared with manual results (p-value = 0.04 and p-value 0.09 when compared with manual and automatic methods respectively, Mann-Whitney U test). The reason for our method's mismatch in segmentation and volumetric performance may be driven by the optimized loss function. As explained in \cite{bertels2019optimization}, soft Dice loss can lead to volumetric bias.

\section{Discussion and Conclusion}
We present a neural network that regresses the AIF for CT perfusion analysis. Unlike previous methods that aim to imitate the AIF selection of a human rater, the training of our network requires no manual annotations. To this end, we implement a differentiable SVD deconvolution, allowing the AIF generating network to be optimized for generating the most discriminative rCBF maps with reference to DWI images.

There are no previous studies that use CNNs to regress the AIF. Moreover, we are first in applying SVD deconvolution differentiability for perfusion applications. Unlike previous works, our approach does not require manual annotations. This is a crucial finding for devising automatic methods free from manual rater's influence. From a scientific point, it is interesting that our approach generates the `best' AIF: current guidelines for the manual selection of AIFs do not have that guarantee.

Our experiments with the ISLES18 data yielded results matching expert segmentation performance. The rCBF maps that we obtained are slightly more informative for finding core-lesions than the ones an expert generates, as shown in the ROC analysis. In all segmentation metrics considered our method is comparable to an expert or an FDA-cleared software.

In future work, we aim to work with extra datasets and incorporate additional perfusion parameters. We currently only have 94 subjects and since each subject corresponds to a single regression, we effectively have only 94 samples. We expect that a larger dataset will further improve results. Similarly, additional data is needed to increase confidence in the method's performance.

This work only optimizes the rCBF map. In future work, we also intend to optimize the Tmax parameter map (which is defined as $argmax_t k(t))$. Tmax estimation is crucial for the applicability of the method and for improving the lesion quantification since an increased Tmax is indicative for tissue at risk. This will require finding a differentiable substitute for the $argmax$ and a ground truth for tissue at risk (e.g. the final infarct in patients that did not have reperfusion).
\subsubsection{Acknowledgements}
This project received funding from the European Union's Horizon 2020 research and innovation program under the Marie Sklodowska-Curie grant agreement TRABIT No 765148. EDLR, DR and DMS are employees of ico\textbf{metrix}. DR is supported by an innovation mandate of Flanders Innovation \& Entrepreneurship (VLAIO).

\bibliographystyle{splncs04}
\bibliography{mybibfile}
%
% \begin{thebibliography}{8}
% \bibliography{mybibfile}

% \end{thebibliography}
\end{document}